\documentclass[10pt, conference, twocolumn]{IEEEtran}
\usepackage{cite}      
\usepackage{siunitx}
\usepackage{array}
\usepackage{amssymb,amsmath}
\usepackage{comment}
\usepackage{array}
\usepackage{scalefnt}
\usepackage[demo]{graphicx}
\usepackage{caption}
\usepackage{subcaption}
\usepackage{flushend}
\usepackage{tikz}
\usepackage{circuitikz}
\usetikzlibrary{shapes,decorations,shadows}
\usetikzlibrary{decorations.pathmorphing}
\usetikzlibrary{decorations.shapes}
\usetikzlibrary{fadings}
\usetikzlibrary{patterns}
\usetikzlibrary{calc}
\usetikzlibrary{decorations.text}
\usetikzlibrary{decorations.footprints}
\usetikzlibrary{decorations.fractals}
\usetikzlibrary{shapes.gates.logic.IEC}
\usetikzlibrary{shapes.gates.logic.US}
\usetikzlibrary{fit,chains}
\usetikzlibrary{positioning}
\usepgflibrary{shapes}
\usetikzlibrary{scopes}
\usepackage{comment}
\usepackage{pgfplots}
\usepackage{pgfplotstable}

\pgfplotsset{
    /pgfplots/flexible xticklabels from table/.code n args={3}{%
        \pgfplotstableread[#3]{#1}\coordinate@table
        \pgfplotstablegetcolumn{#2}\of{\coordinate@table}\to\pgfplots@xticklabels
        \let\pgfplots@xticklabel=\pgfplots@user@ticklabel@list@x
        \pgfkeys{/pgf/fpu=true}
    }
}

\newcommand{\plotwidth}{\linewidth}
\newcommand{\plotheight}{0.9\linewidth}

\pgfplotsset{
    discard if not/.style 2 args={
        x filter/.code={
            \edef\tempa{\thisrow{#1}}
            \edef\tempb{#2}
            \ifx\tempa\tempb
            \else
                
            \fi
        }
    },
}

\tikzset{min-sum 5 style/.style={color=blue,dashed,mark=+,mark options={solid,scale=2}}}

\tikzset{min-sum 10 style/.style={color=blue,dashed,mark=o,mark options={solid,scale=1.2}}}

\tikzset{min-sum 100 style/.style=
    	{color=blue,dashed,mark=x,mark options={solid,scale=2}}}
  
\tikzset{BP style/.style=
    	{color=black,dashed,mark=triangle}}

\author{\IEEEauthorblockN{Tasnuva Tithi, Chris Winstead and Gopalakrishnan Sundararajan\\} \IEEEauthorblockA{Department of Electrical And Computer Engineering\\ Utah State University\\Logan, UT 84322-4120\\Email: tasnuvatithi@aggiemail.usu.edu, chris.winstead@usu.edu, gopal.sundar@aggiemail.usu.edu}}

\title{Decoding LDPC codes via Noisy Gradient Descent Bit-Flipping with Re-Decoding}

\begin{document}
\maketitle

\begin{abstract}
In this paper, we consider the performance of the Noisy Gradient Descent Bit Flipping (NGDBF) algorithm under re-decoding of failed frames. NGDBF is a recent algorithm that uses a non-deterministic gradient descent search to decode low-density parity check (LDPC) codes. The proposed re-decode procedure obtains improved performance because the perturbations are independent at each re-decoding phase, therefore increasing the likelihood of successful decoding. We examine the benefits of re-decoding for an LDPC code from the IEEE 802.3an standard, and find that only a small fraction of re-decoded frames are needed to obtain significant performance benefits. When re-decoding is used, the NGDBF performance is very close to a benchmark offset min-sum decoder for the 802.3an code. 
\end{abstract}

\section{Introduction}

A considerable amount of research has been done in the past decade to reduce the complexity of soft-decision algorithms for Low Density Parity Check (LDPC) codes.  The class of Weighted Bit Flipping algorithms (WBF) provide a good trade-off between performance and complexity \cite{Kou_2000}. The WBF algorithms employ an inversion function at the symbol nodes to decide on the bits that needed to be flipped during a decoding iteration. Many variants to the WBF algorithms have been proposed \cite{Wu_2007b} \cite{Zhang_2004}. One such variant is Gradient descent bit-flipping (GDBF) algorithm \cite{Wadayama_2010_TCOMM}. The GDBF algorithm formulates the decoding problem as a gradient descent search problem and provides good performance with very low complexity. The major drawback of the GDBF algorithms is that they tend to be trapped in spurious local maxima. In the context of LDPC decoding, local maxima can be interpreted as trapping sets which prevent the decoder from converging onto a valid codeword \cite{Richardson_2003_ACCC}.

The recently proposed Noisy GDBF (NGDBF) algorithm provides superior performance by injecting random noise into the GDBF inversion function \cite{Sundararajan_2014_TCOMM}. The noise perturbation is hypothesized to disrupt the activity around trapping sets, so that the decoder has a chance to escape. In this work, we observe that failed NGDBF frames can often be correctly decoded by re-decoding the algorithm from its initial state. Re-decoding is shown through simulations to be more effective than increasing the iterations. To explain this result, we hypothesize that the noise perturbations may sometimes stimulate new trapping set conditions that are less likely to be escaped. Because the NGDBF algorithm is non-deterministic, these conditions will not necessarily recur when decoding is performed a second time.

Redecoding has been previously examined for non-deterministic decoding algorithms \cite{sharifi2010majority, Leduc_2012, Cushon_2014}. Tehrani et al.\ \cite{sharifi2010majority} speculated that re-decoding  provides an opportunity for a stochastic decoder to evade the dominant (8,\,8) trapping set in the 802.3an LDPC code. In this work, we examine the same hypothesis applied to the NGDBF algorithm. 

The remainder of this paper is organized as follows: Section II reviews the GDBF and NGDBF algorithms. Section III describes the proposed re-decoding method and Section IV details the simulation results. Conclusions are drawn in Section V.

\section{Background}
\subsection{Notation:}
We denote our parity check matrix by $H$, which is a binary matrix with dimensions $m \times n$; $n>m \geq 1$; $\lbrace m, n\rbrace \in \mathbb{Z^{+}}$. $H$ is a sparse matrix, and $H_{ij}=1$, corresponds to a connection between the $i\textsuperscript{th}$ parity check node and $j\textsuperscript{th}$ symbol node; $i \in [1 ~~m]$, $j \in [1~~n]$. The set of codewords associated to $H$ is $C \triangleq \lbrace \vec{c} \in \lbrace 0,1\rbrace^n: Hc=0\rbrace$. For AWGN transmission, $C$ is mapped to $\hat{C} \triangleq \hat{c} \in \lbrace {1,-1} \rbrace^n$. The AWGN transmission is defined by $\vec{y}=\hat{c}+\vec{z}$. Here, $\vec{y}$ is the received vector at the receiver, and $\vec{z}$ is an $n$ dimensional vector of independent and identically distributed Gaussian random noise; with zero mean and variance = $N_0/2$, where $N_0$ is the noise spectral density.

The neighborhood of a node refers to all the other nodes that are connected to it. In this paper, we denote the neighborhood of the $i\textsuperscript{th}$ check node by $\mathcal{N}(i) \triangleq \lbrace j: h_{ij}=1 \rbrace$, and the neighborhood of the $j\textsuperscript{th}$ symbol node by $\mathcal{M}(j)\triangleq \lbrace i: h_{ij}=1 \rbrace$; where $h_{ij}$ is an element of the parity check matrix $H$.

The decision vector at each iteration is denoted by $\vec{x} \in \lbrace -1,1 \rbrace ^n$. If $\vec{x}$ is a valid codeword and $\vec{x} \in \hat{C}$, all the parity checks would be satisfied. The $i\textsuperscript{th}$ parity check operation is denoted by $s_i \triangleq \prod_{j\in \mathcal{N}(i)} x_j$. Therefore, a satisfied parity check equation refers to $s_i= 1$. If at iteration $t$, all the parity checks are satisfied, then $\vec{x}(t)$ is declared as a valid codeword and decoding is terminated. In all algorithms, unsuccessful frames are terminated after a maximum number of iterations $T$.

\subsection{The GDBF algorithm and NGDBF algorithms}
Wadayama et al.\ proposed the GDBF algorithm for decoding LDPC codes \cite{Wadayama_2010_TCOMM}. Several variants of GDBF have been described; in this work we consider only the fully-parallel multi-bit GDBF algorithm.
%
%
The NGDBF algorithm, described by Sundararajan et al.\  \cite{Sundararajan_2014_TCOMM}, modifies GDBF by adding a pseudo-random perturbation to the inversion function at each symbol node during each iteration. The perturbations consist of identical and independently distributed Gaussian random noise samples $q_k$ with zero mean and variance equal to $\sigma^2= \eta^2 N_0/2$ --- proportional to the variance of the channel noise --- where $\eta\geq 0$ is a noise scale parameter. The resulting inversion function is 
$$E_{k} = x_{k}y_{k} + w_{k}\sum_{i\in \mathcal{M}\left(k\right)} s_i + q_{k},$$
which describes GDBF when $\eta = 0$ and $w=1$. For each symbol node index $k$, the corresponding decision $x_k$ is flipped if $E_k < \theta$, where $\theta\in\mathbb{R}^-$ is an empirically determined threshold value.

In order to obtain the best performance, NGDBF makes use of additional heuristics. The heuristics used in this work include output smoothing, in which a sliding-window average is taken on the output decisions, and a threshold adaptation procedure devised by Ismail et al.\ \cite{Ismail_2010}. With adaptive thresholding, each symbol has a local threshold $\theta_k$ which is adjusted dynamically. In \cite{Sundararajan_2014_TCOMM}, this combination of heuristics is referred to as SM-NGDBF. Throughout this paper, any references to NGDBF shall refer to the algorithm described as follows:

\begin{enumerate}[\setlabelwidth{Step 3}]

\item[\textbf{Step 0:}] Initialize $\theta_k\left(t=0\right) = \theta$ for all $k$, where $\theta$ is the global initial threshold parameter. Optionally saturate sample magnitudes at $y_{\rm max}$ and set $\vec{x}={\rm sign}\left(\vec{y}\right)$.

\item[\textbf{Step 1:}] Compute syndrome components: 
	\begin{equation}
		s_i = \prod_{j\in \mathcal{N}\left(i\right)}x_{j},
	\end{equation}
for all $i \in \left\{1, 2,...., m\right\}$. If $s_{i}=+1$ for all $i$, output $x$ and stop.

\item[\textbf{Step 2:}] Compute inversion functions: 
	\begin{equation}
		E_{k} = x_{k}y_{k} + w_{k}\sum_{i\in \mathcal{M}\left(k\right)} s_i + q_{k} \label{eq:ngdbfinvfun1} 
	\end{equation}
for $k \in \{1,\,2,\,\dots,\,n\}$. 
where $w_{k}$ is a syndrome weight parameter and $q_{k}$ is a Gaussian distributed random variable with zero mean and variance $\sigma^2 = \eta^2 N_{0}/2$, where $0 < \eta \leq 1$.  All $q_k$ are independent and identically distributed.

\item[\textbf{Step 3:}]Bit-flip operations:    if $E_k\left(t\right) < \theta_k\left(t\right)$ then $x_k\left(t+1\right) = -x_k\left(t\right)$,
    otherwise $\theta_k\left(t+1\right) = \lambda\theta_k\left(t\right),$
%
where $\lambda$ is a global threshold adaptation parameter for which $0 < \lambda \leq 1$.

\item[\textbf{Step 4:}] Repeat steps 1 to 3 until all $s_i=1$, for $i=1,\,2,\,\dots,\,m$, or maximum number of iterations $T$ is reached.
\end{enumerate}

\noindent In this work we make use of the smoothing heuristic where an up/down counter is placed at the output of each symbol. Each counter is initialized at zero, and updated as follows:
\begin{equation}
X_k(t+1)=X_k(t)+x_k(t)
\label{eq:ngdbfsmoothing}
\end{equation}
If all the parity checks are satisfied by $x_k$, then $x_k$ is declared as the final result; otherwise the smoothed decision is used, $\bar{x}_k=\textrm{sign}(X_k)$. 

\section{Re-decoding and Trapping Sets}

Re-decoding from the same initial condition was considered previously for stochastic decoders by Tehrani et al.\  \cite{sharifi2010majority}, who proposed it as a method to evade trapping sets. In that paper, the authors demonstrated BER improvement due to re-decoding, but did not provide a detailed inspection of trapping set behavior. In this section, we present some experiments on  the (8,8) absorbing set known to be dominant in the 802.3an 10GBASE-T standard LDPC code under belief propagation \cite{Zhang_Schlegel_2013}. The induced graph for this set is shown in Fig.~\ref{fig:trap8_8}, where the degree-one check nodes are indicated as \tikz{\node[rectangle, inner sep=0, draw=black, fill=black, minimum width=0.2cm,minimum height=0.2cm] at (0,0) {};}, degree-two check nodes as \tikz{\node[rectangle, inner sep=0, draw=black, fill=white, minimum width=0.2cm,minimum height=0.2cm] at (0,0) {};}, and symbol nodes as \tikz{\node[circle, inner sep=0, draw=black, fill=black, minimum width=0.2cm] at (0,0) {};}. While a full trapping set analysis has not yet been developed for NGDBF, in this section we verify that the (8,\,8) set acts as a trapping set for GDBF and NGDBF, and we inspect the dynamics that allow NGDBF to evade the trapping set during decoding. 

To investigate NGDBF dynamics on this absorbing set, a localized simulation was performed on the (8,\,8) subgraph. The correct state is assumed to be $\hat{c}=\left(+1\,+1\,\dots +1\right)$. The GDBF and NGDBF algorithms were simulated with identical inputs $\vec{y}=\hat{c}+\vec{z}$, where $\vec{z}$ is a vector of zero-mean Gaussian noise samples with $\sigma=1$. The simulations were performed with parameters $\lambda=1$ (i.e.\ without threshold adaptation), $w=1$, $T=100$ and for NGDBF $\eta=1$. In these simulations, a frame was considered successful if the correct result, $\vec{x}=\hat{c}$, was obtained for at least one iteration. Failed frames were saved for detailed inspection.

It was found that failed frames typically begin in a metastable initial condition, where one or two early flips determine the ultimate trajectory. Fig.~\ref{fig:NGDBF_escape} shows the trajectory of inversion functions for a case in which GDBF becomes trapped in an oscillating cycle, but NGDBF avoids the oscillation due to a fortuitous early flip. In the early iterations, some of the $E_k$ are negative or weakly positive, so they are likely to be flipped. In later iterations, most of the NGDBF $E_k$ values are strongly positive, so additional flips are unlikely in spite of the noise perturbations.

\begin{figure}
\centering
\begin{tikzpicture}
  \def\idx{0}
  \begin{axis}[
    width=0.9\linewidth,height=0.75\linewidth,
    xmin=0,xmax=20,
    ymin=-5,ymax=15,
    xlabel=Iteration,
    ylabel=$E_k$,
    legend entries={NGDBF,GDBF},
    ]
    \addlegendimage{no markers,black}
    \addlegendimage{no markers,blue,dashed}
    \foreach \x in {0,1,...,7}
    \addplot[no marks,black,forget plot] table[x expr=\coordindex, y index=\x] {data/replay\idx/E_gdbf_fail_replay\idx.dat};

    \foreach \x in {0,1,...,7}
    \addplot[no marks,blue,dashed,forget plot] table[x expr=\coordindex, y index=\x] {data/replay\idx/E_ngdbf_fail_replay\idx.dat};
   \end{axis}
\end{tikzpicture}
\caption{A case where NGDBF settles on an all-error pattern on the (8,\,8) absorbing set. Error propagation is triggered by a single errant bit-flip that occurs in the first five iterations. }
\label{fig:NGDBF_fail}
\end{figure}
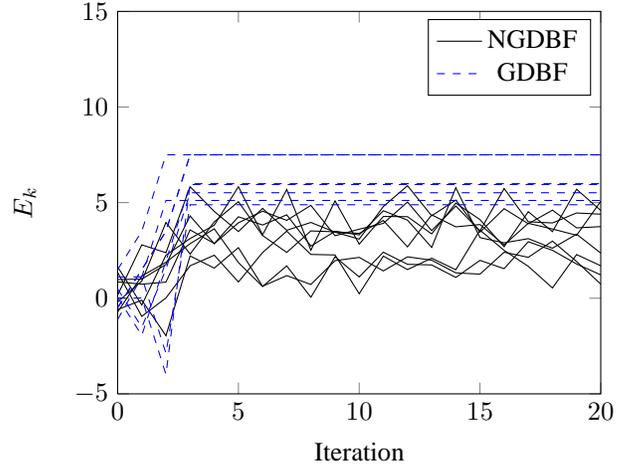

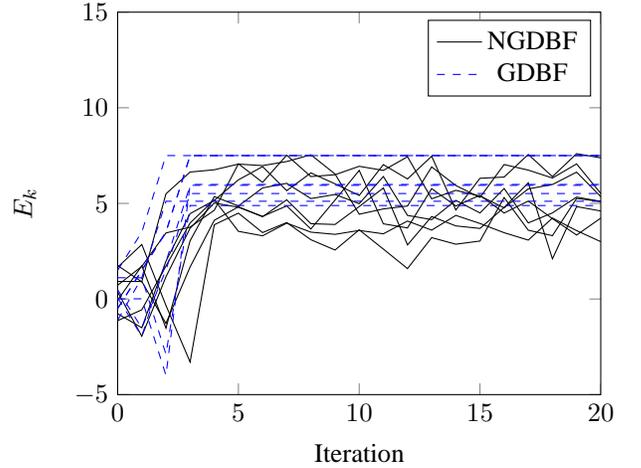
\begin{figure}
\centering
\begin{tikzpicture}
  \def\idx{1}
  \begin{axis}[
    width=0.9\linewidth,height=0.75\linewidth,
    xmin=0,xmax=20,
    ymin=-5,ymax=15,
    xlabel=Iteration,
    ylabel=$E_k$,
    legend entries={NGDBF,GDBF},
    ]
    \addlegendimage{no markers,black}
    \addlegendimage{no markers,blue,dashed}

    \foreach \x in {0,1,...,7}
    \addplot[no marks,black] table[x expr=\coordindex, y index=\x] {data/replay\idx/E_gdbf_fail_replay\idx.dat};
    
    \foreach \x in {0,1,...,7}
    \addplot[no marks,blue,dashed] table[x expr=\coordindex, y index=\x] {data/replay\idx/E_ngdbf_fail_replay\idx.dat};
   \end{axis}
\end{tikzpicture}\caption{A re-decoded case with the same initial conditions as Fig.~\ref{fig:NGDBF_fail}. This time NGDBF evades the erroneous state and corrects all errors.}
\label{fig:NGDBF_replay}
\end{figure}

Fig.~\ref{fig:NGDBF_fail} shows a case where NGDBF failed due to an errant early flip. In this case, NGDBF eventually converged on an all-error state with positive $E_k$. Because the $E_k$ are positive, future flips are unlikely to occur and the error state is effectively stable. Fig.~\ref{fig:NGDBF_replay} shows a repeated simulation from the same initial condition. In this case, NGDBF made different flips in the first five iterations, and converged on the correct state. This example demonstrates advantages of re-decoding from the same initial state, which cannot be achieved by extending the simulation time. The benefits of re-decoding are more pronounced when threshold adaptation is used, since the evolving thresholds tend to harden the stability of the final state, thereby lowering the probability that NGDBF will escape to the correct state if given more iterations. 

These experiments were repeated for several values of $\sigma$, and NGDBF was found to have a consistently lower rate of converging on an erroneous state compared to GDBF. The simulation method used here is illuminating about the dynamics and provides motivation for re-decoding, but it is not sufficient to quantify the frame error probability associated with this absorbing set. For $\sigma<0.7$, we did not obtain any failed cases for NGDBF. In the sequel, we evaluate the re-decoding method for two practical codes, and show that significant performance benefits are obtained. 

%

\begin{figure}
\centering
\begin{tikzpicture}[scale=1.25,transform shape]
  \foreach \x in {1, 2, ..., 8} {
    \node[circle, inner sep=0, draw=black, fill=black, minimum width=0.25cm] at ({\x*45}:2cm) (n\x) {};
    \node[rectangle,draw=black,fill=black,minimum width=0.1cm, minimum height=0.1cm, inner sep=0] at ({\x*45}:2.5cm) (p\x) {};
    \path (n\x) edge (p\x);
   }

  \foreach [count=\r] \row in {{0,1,1,1,0,0,1,1},
     {0,0,1,1,0,0,1,1},
     {0,0,0,1,1,1,0,0},
     {0,0,0,0,1,1,0,0},
     {0,0,0,0,0,1,1,1},
     {0,0,0,0,0,0,1,1},
     {0,0,0,0,0,0,0,1},
     {0,0,0,0,0,0,0,0}}{
        \foreach [count=\c] \cell in \row{
            \ifnum\cell=1%
                \path[every node/.style={sloped,anchor=base,rectangle,draw=black,fill=white,auto=false,minimum width=0.1cm, minimum height=0.1cm, inner sep=0}] 
                (n\r) edge node {} (n\c);
            \fi
        }
    }

\end{tikzpicture}
\caption{The dominant (8,\,8) absorbing set in the 802.3an 10GBASE-T LDPC code.}
\label{fig:trap8_8}
\end{figure}
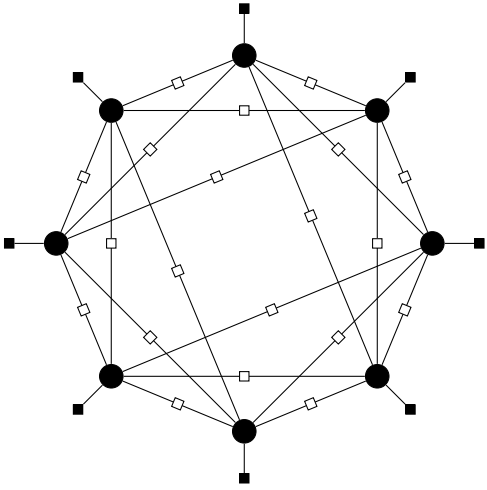

\begin{figure}
\centering
\begin{tikzpicture}
  \begin{axis}[
    width=0.9\linewidth,height=0.75\linewidth,
    xmin=0,xmax=20,
    ymin=-5,ymax=15,
    xlabel=Iteration,
    ylabel=$E_k$,
    legend entries={NGDBF,GDBF},
    ]
    \addlegendimage{no markers,black}
    \addlegendimage{no markers,blue,dashed}
    \foreach \x in {0,1,...,7}
    \addplot[no marks,blue,dashed,forget plot] table[x expr=\coordindex, y index=\x] {data/fail/E_ngdbf_fail_good.dat};

    \foreach \x in {0,1,...,7}
    \addplot[no marks,black,forget plot] table[x expr=\coordindex, y index=\x] {data/fail/E_gdbf_fail_good.dat};

   \end{axis}
\end{tikzpicture}
\caption{A typical case where GDBF is trapped but NGDBF escapes due to random perturbations.}
\label{fig:NGDBF_escape}
\end{figure}
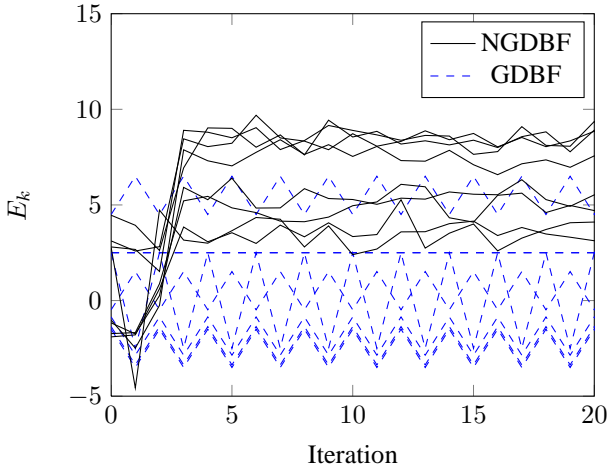

\section{Simulation results}

Simulations were performed using re-decoding with NGDBF for two codes: the rate $1/2$ regular (3,\,6) LDPC code identified as PEGReg504x1008 in MacKay's online encyclopedia of sparse graph codes \cite{Mackay_2013:Online} and the rate $0.8413$ LDPC code defined in IEEE 802.3 standard. The smoothing and threshold adaptation heuristics are used for the PEGReg504x1008 code, and the algorithm name is indicated as SM-NGDBF in reported simulations. For the 802.3 code, these heuristics were not used. Each frame was allowed to be simulated up to a maximum number $\Phi$ of re-decoding phases. At least 200 bit errors and at least 20 word errors were observed to obtain the BER measurements for each simulation. 

\begin{figure}
\begin{tikzpicture}
\begin{semilogyaxis}[
    width=\plotwidth,
    height=\plotheight,
    xlabel=$E_{b}/N_{0}$ (dB),
    ymax = 0.1,  
    ylabel=BER, 
    grid=both, 
    legend style={
        legend columns=1,
        legend cell align=center,
        font=\tiny},
	]											

			
\addplot [red, solid, mark=diamond*] table [x index=0, y index=1]  {data/averaged_ber.txt};
\addlegendentry{SM-NGDBF ($\Phi=10$)};

\addplot+[min-sum 5 style,discard if not={T}{5}] table[x=SNR,y=BER] {data/NMS_best.tab};
\addlegendentry{NMS ($T=5$)};
			
\addplot+[min-sum 10 style,discard if not={T}{10}] table[x=SNR,y=BER] {data/NMS_best.tab};
\addlegendentry{NMS ($T=10$)};

\addplot+[min-sum 100 style,discard if not={T}{100}] table[x=SNR,y=BER] {data/NMS_best.tab};
\addlegendentry{NMS ($T=100$)};

\addplot+[BP style] table[x=SNR,y=BER] {data/bp.tab};
\addlegendentry{BP};
	
\end{semilogyaxis} 
\end{tikzpicture}
\caption{BER results for the PEGReg504x1008 code. Results for belief propagation (BP) and normalized min-sum (NMS) with different iterations are provided for comparison.}
\label{fig:BERpegreg}  
\end{figure}
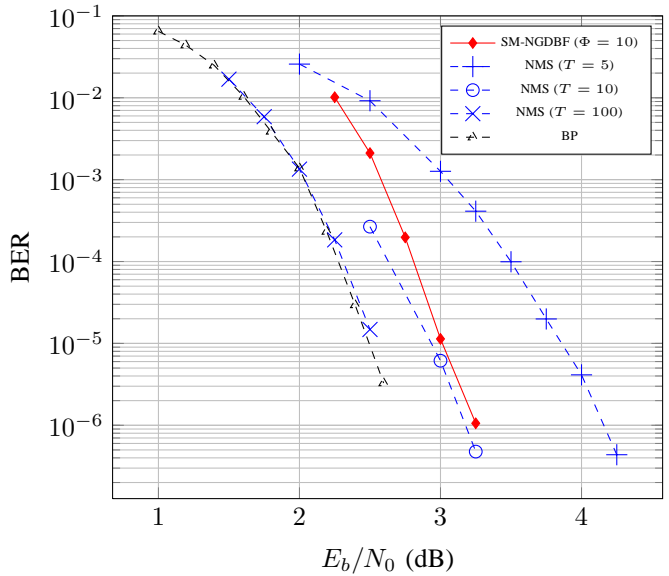

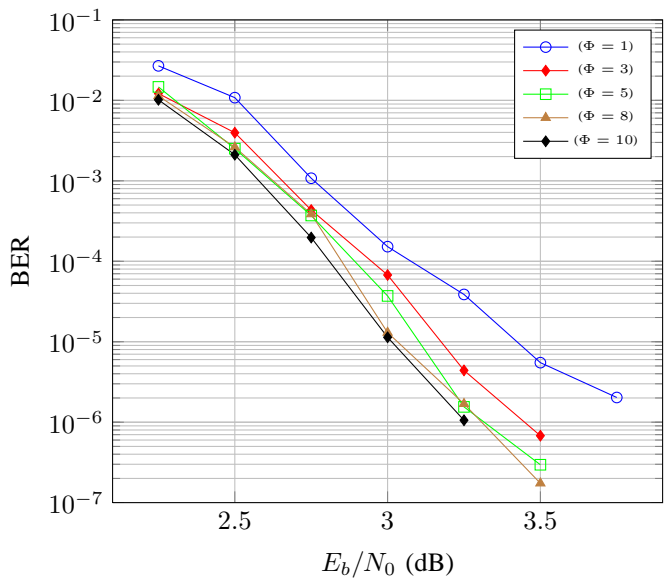
\begin{figure}
\begin{tikzpicture}
\begin{semilogyaxis}[
    width=\plotwidth,
    height=\plotheight,
    xlabel=$E_{b}/N_{0}$ (dB),
    ymax = 0.1,
    ymin=1e-7,  
    ylabel=BER, 
    grid=both, 
    legend style={
        legend columns=1,
        legend cell align=center,
        font=\tiny},
	cycle list={{blue,solid,mark=o},{red,solid,mark=diamond*},{green,solid, mark=square},{brown,solid,mark=triangle*},{black, solid,mark=diamond*}} 	] 
												
\addplot table [x index=0, y index=1]  {data/smngdbf_results.txt};
\addlegendentry{($\Phi=1$)};

\addplot table[x index=0, y index=2]  {data/all_the_phases.txt};
\addlegendentry{($\Phi=3$)};

\addplot table[x index=0, y index=4]  {data/all_the_phases.txt};
\addlegendentry{($\Phi=5$)};

\addplot table[x index=0, y index=7]  {data/all_the_phases.txt};
\addlegendentry{($\Phi=8$)};

\addplot table[x index=0, y index=1]  {data/averaged_ber.txt};
\addlegendentry{($\Phi=10$)};

\end{semilogyaxis} 
\end{tikzpicture}
\caption{BER for re-decoding with the SM-NGDBF on the PEGReg504x1008 code for different $\Phi$s.}
\label{fig:Phases}  
\end{figure}

For the PEGReg504x1008 code, SM-NGDBF was simulated with parameters $T=300$, $w = 0.816$, $\lambda= 0.98$, initial threshold $\theta = -0.6$ and noise scale $\eta= 0.75$. Fig.~\ref{fig:BERpegreg} shows BER performance results for the PEGReg504x1008 code with $\Phi=10$. The re-decoding technique has significant gain when applied to the SM-NGDBF algorithm. At BER$=10^{-6}$, re-decoding provides a gain of about 0.5 dB. Output smoothing is only used for iterations exceeding $(T-64)$. Performance of NMS and BP algorithms are presented for comparison. 
The BER improves with higher values of $\Phi$, but there is a diminishing benefit as $\Phi$ is increased. As seen in Fig.~\ref{fig:Phases}, there is a rapid improvement in performance from $\Phi= 1$ through $\Phi=5$. As $\Phi$ is increased further, the improvement in BER performance becomes less significant. There is slight improvement in $\Phi=10$ compared to the improvement in the earlier phases. 

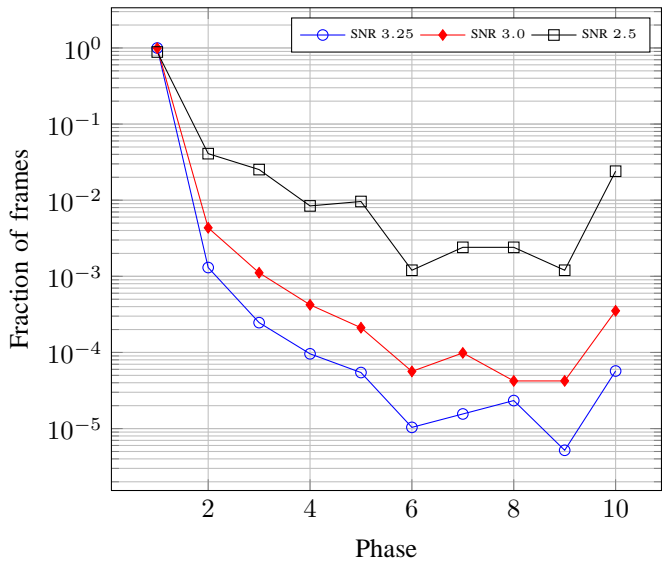
\begin{figure}
\begin{tikzpicture}
  \begin{semilogyaxis}[width=\plotwidth,
    height=\plotheight,
    xlabel=Phase,
    ylabel= Fraction of frames,
    grid=both,
    legend style={
	legend columns= 3,
	legend cell align=center,
	font=\tiny},
	cycle list={{blue,solid,mark=o},{red,solid,mark=diamond*},{black,solid, mark=square},{brown,solid,mark=triangle*}
			}
		]
		\addplot table[x index=0, y index=1] {data/phase_hist.txt};
		\addlegendentry{SNR $3.25$};
		\addplot table[x index=0, y index=2] {data/phase_hist.txt};
		\addlegendentry{SNR $3.0$};
		\addplot table[x index=0, y index=3] {data/phase_hist.txt};
		\addlegendentry{SNR $2.5$};
		
\end{semilogyaxis}
\end{tikzpicture}
\caption{A histogram showing the fraction of frames completed at each decoding phase for SM-NGDBF on the PEGReg504x1008 code, $\Phi= 10$. The increase of frames at the last phase arises due to the accumulation of failed frames.}
\label{fig:phaseplot}
\end{figure}

The simulations show that re-decoding is necessary for a small fraction of frames. Fig.~\ref{fig:phaseplot} shows the distribution of re-decoding phases to complete decoding. Most of the failed frames that are not corrected in the first decoding phase are corrected by the second phase. The frames that are not corrected by the second phase are passed onto the third phase and so forth. Finally, at the last phase, the accumulated failed frames determine the word error rate (WER). This accumulation is evident in the last phase in Fig.~\ref{fig:phaseplot}. 

\begin{figure}
\begin{tikzpicture}

\begin{semilogyaxis}[
    width=\plotwidth,
    height=\plotheight,
    xlabel=$E_{b}/N_{0}$ (dB),
    ymax = 0.1,  
    ymin = 1e-7,
    xmax = 4.5,
    xmin = 3,
    ylabel=BER, 
    grid=both, 
    legend style={
        legend cell align=left,
        font=\tiny},
	cycle list={{blue,solid,mark=o},{red,solid,mark=*},{black,solid, mark=square},{brown,solid,mark=square*}}
    ] 

  \addplot+[discard if not={MAXPHASE}{1}] table[x expr=\thisrow{SNR}-0.1513195549,y=BER] {data/redecode_MNGDBF_09_09_2014.dat};
  \addlegendentry{NGDBF};



  \addplot+[discard if not={MAXPHASE}{1}] table[x expr=\thisrow{SNR}-0.1513195549,y=BER] {data/redecode_MNGDBF_09_09_2014.dat};
  \addlegendentry{NGDBF ($\Phi=1$)};

  \addplot+[discard if not={MAXPHASE}{2}] table[x expr=\thisrow{SNR}-0.1513195549,y=BER] {data/redecode_MNGDBF_09_09_2014.dat};
  \addlegendentry{NGDBF ($\Phi=2$)};

  \addplot+[discard if not={MAXPHASE}{8}] table[x expr=\thisrow{SNR}-0.1513195549,y=BER] {data/redecode_MNGDBF_09_09_2014.dat};
  \addlegendentry{NGDBF ($\Phi=8$)};
 

  \addplot [black, dashed, mark=triangle*] table[x expr=\thisrow{SNR},y=BER] {data/OMS_802_3.dat};
  \addlegendentry{OMS ($T=20$)};

  \addplot [black, dashed, mark=x] table[x=SNR,y=BER] {data/Zhang.dat};
  \addlegendentry{OMS ($T=8$)};
\end{semilogyaxis}
\end{tikzpicture}
 \caption{ BER for re-decoded NGDBF compared to a benchmark OMS decoder for the IEEE 802.3 standard LDPC code.}
\label{fig:802_3_performance}
\end{figure}
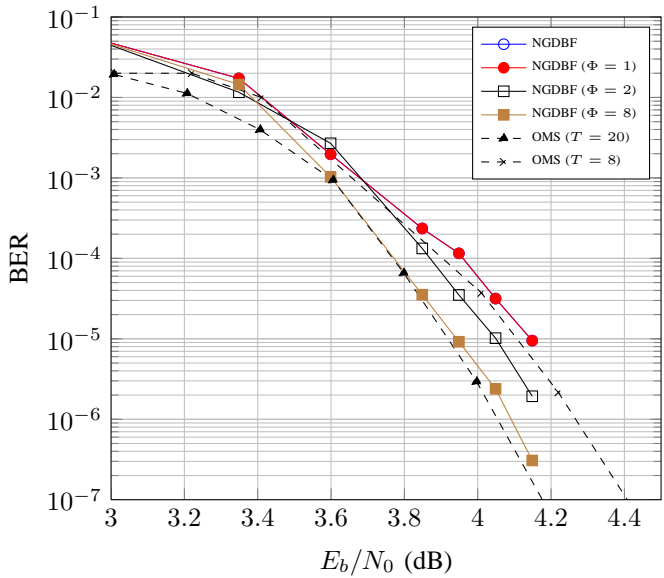

Fig.~\ref{fig:802_3_performance} shows the BER performance results for NGDBF on the IEEE 802.3 standard LDPC code. Smoothing is not used for simulations with this code, because no significant improvement was achieved using smoothing in this case. The simulation parameters are $T= 1000$, $w=0.20833$, $\lambda= 1$, $\theta = -0.525$, and $\eta= 0.92$. Since $\lambda=1$, threshold adaptation is not required for this code. To evaluate the performance we use a recently reported 802.3 Offset Min-Sum (OMS) decoder as a benchmark \cite{Zhang_2010a}. Re-decoding provides a gain of 0.25 dB  for this code. The BER performance of NGDBF is very close to the benchmark OMS decoder.

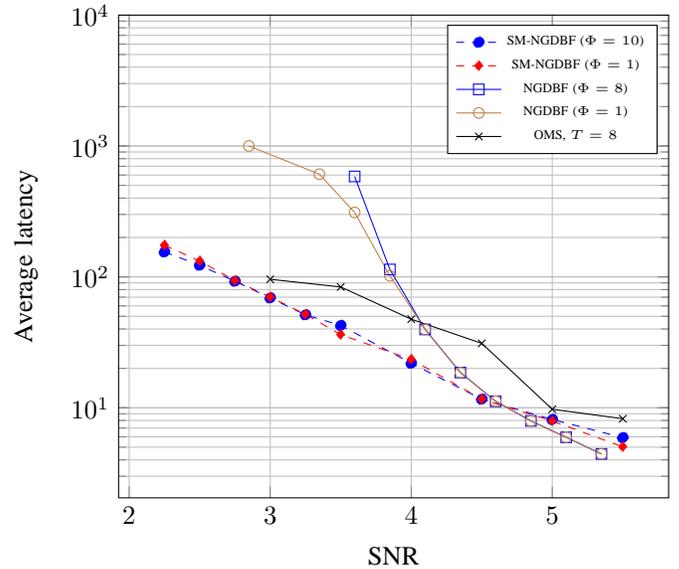
\begin{figure}
\begin{tikzpicture}
\begin{semilogyaxis}[width= \plotwidth,
height=\plotheight,
ymax=10000,
xlabel=SNR,
ylabel=Average latency,
grid=both,
   legend style={
	legend columns=1,
	legend cell align=center,
	font=\tiny},
	cycle list={{blue,dashed,mark=*},{red,dashed,mark=diamond*},{blue,solid, mark=square},{brown,solid,mark=o},{black,solid,mark=diamond}, {red,solid, mark =triangle}
			}
		]
\addplot table[x index=0, y index=1] {data/averaged_iteration_pegreg.txt};
\addlegendentry{SM-NGDBF ($\Phi=10$)};

\addplot table[x index=0, y index=2] {data/averaged_iteration_pegreg.txt};
\addlegendentry{SM-NGDBF ($\Phi=1$)};

\addplot+[discard if not={MAXPHASE}{8}] table[x expr=\thisrow{SNR}-0.1513195549,y=ITER] {data/iterations_MNGDBF_09_09_2014.dat};
\addlegendentry{NGDBF ($\Phi=8$)};
	
\addplot+[discard if not={MAXPHASE}{1}] table[x expr=\thisrow{SNR}-0.1513195549,y=ITER] {data/iterations_MNGDBF_09_09_2014.dat};
\addlegendentry{NGDBF ($\Phi=1$)};

\addplot [black,solid,mark=x] table[x index=0, y index=2] {data/Zhang_iterations.txt};
\addlegendentry{OMS, $T=8$};				

\end{semilogyaxis}
\end{tikzpicture}
\caption{Latency comparison between different algorithms and codes. The dashed lines indicate simulations on the  PEGReg504x1008 code, and solid lines indicate simulations on the IEEE 802.3 standard LDPC code. 
}
\label{fig:latency}
\end{figure}

Since re-decoding incurs a substantial latency penalty, we examined the average latency on the 802.3 code. If buffering can be tolerated by the end application, then the average latency penalty is very small since nearly all frames are successfully decoded in the first phase. Fig.~\ref{fig:latency} shows the average latency in terms of clock cycles for all simulated cases, in comparison to the reported latency of the OMS decoder. The OMS decoder uses a semi parallel layered architecture which requires 12 clock cycles to complete an iteration. The reported average number of iterations is therefore scaled by 12 to obtain the average latency depicted in Fig.~\ref{fig:latency}. The NGDBF decoder has a much lower complexity than the OMS algorithm, so layering is not required and we can expect every iteration to complete in a single clock cycle. The number of iterations is therefore equivalent to the latency in the NGDBF case. Fig.~\ref{fig:latency} shows that the average latency for NGDBF is quite large at low SNR values, but decreases at higher SNRs where it has a lower latency than OMS. When operating at higher SNR values, re-decoded NGDBF offers better performance and lower average latency than the benchmark design.  To account for consecutive worst case frames, the NGDBF decoder would require a larger frame buffer compared to the OMS decoder. This is a potential drawback of re-decoding with the NGDBF algorithm.

\section{Conclusion}
The NGDBF algorithm has been modified by the application of the re-decoding method described in this paper. NGDBF is a low complexity algorithm compared to stochastic decoding, OMS, NMS or BP algorithms. Re-decoding takes advantage of the inherent random nature of NGDBF and further enhances its performance. Re-decoding provides a gain of up to 0.5 dB over the original NGDBF algorithm for the codes examined in this paper. For the IEEE 802.3 standard LDPC code, re-decoding yields performance very close to a benchmark OMS decoder design and requires less average latency. 

The potential drawbacks to re-decoding include high latency when operating at low SNRs. Frame buffering is also needed to accommodate the additional delay required for re-decoding unsuccessful frames. The applicability of re-decoding is application dependent, since it may not be possible to delay frame delivery.
The buffering requirement is relaxed at higher SNRs, as a smaller fraction of frames utilize the re-decoding phases. Future research can be done to optimize the buffer size and frame scheduling for consecutive worst case frames at low SNRs. It would also be interesting to quantify absorbing set contributions to the error rate in NGDBF, where re-decoding may offer benefits in the error-floor region. 

\section{Acknowledgment}
\addcontentsline{toc}{section}{Acknowledgment}
This work was supported by the US National Science Foundation under award ECCS-0954747, and by Research Catalyst grant from Utah State University. We also thank Dr. Emmanuel Boutillon and lab-STICC at UBS (Universit\'e de Bretagne Sud), Lorient, France, for their hospitality during the early phase of this work.

\bibliographystyle{IEEEtran}
\bibliography{redecode_ngdbf}

\end{document}